\documentclass[12pt]{iopart}

\usepackage{amssymb}
\usepackage{graphicx}
\usepackage{dcolumn}
\usepackage{bm}
\usepackage{booktabs}
\usepackage{ctable}
\usepackage{color}
\begin{document}

\title[Tat translocation without specific quality control]{Protein translocation without specific quality control in a computational model of the Tat system}
\author{Chitra R. Nayak$^{1}$, Aidan I. Brown$^2$, and Andrew D. Rutenberg$^2$ }

\address{$^1$Department of Physics, University of Toronto, Toronto, ON, Canada, M5S 1A7}
\address{$^2$Department of Physics and Atmospheric Science, Dalhousie University, Halifax, NS, Canada, B3H 4R2}

\ead{\mailto{rchitra.r@gmail.com},\mailto{andrew.rutenberg@dal.ca}}

\begin{abstract}
The twin-arginine translocation (Tat) system transports folded proteins of various sizes across both bacterial and plant thylakoid membranes.   The membrane-associated TatA protein is an essential component of the Tat translocon, and a broad distribution of different sized TatA-clusters is observed in bacterial membranes. We assume that the size dynamics of TatA clusters are affected by substrate binding, unbinding, and translocation to associated TatBC clusters, where clusters with bound translocation substrates favour growth and those without associated substrates favour shrinkage. With a stochastic model of substrate binding and cluster dynamics, we numerically determine the TatA cluster size distribution.  We include a proportion of targeted but non-translocatable (NT) substrates, with the simplifying hypothesis that the substrate translocatability does not directly affect  cluster dynamical rate constants or substrate binding or unbinding rates. This amounts to a translocation model without specific quality control. Nevertheless, NT substrates will remain associated with TatA clusters until unbound and so will affect cluster sizes and translocation rates. We find that the number of larger TatA clusters depends on the NT fraction $f$.  The translocation rate can be optimized by tuning the rate of spontaneous substrate unbinding, $\Gamma_U$. We present an analytically solvable three-state model of substrate translocation without cluster size dynamics that  follows our computed  translocation rates, and that is consistent with {\em in vitro} Tat-translocation data in the presence of NT substrates.
\end{abstract}

\pagebreak
\section{Introduction}

Trafficking of biomolecules across membranes is an essential process for all living cells. The export of proteins is particularly interesting, and there are many distinct mechanisms for protein export from the bacterial cytoplasm \cite{Economou2006}.  Most of these mechanisms export linear peptide chains, which then fold outside the cytoplasm. Remarkably, the twin-arginine translocation (Tat) protein export pathway of bacteria, and the homologous Tat pathway of plant thylakoids, translocates folded proteins --- often with cofactors --- across energized membranes \cite{Palmer2012, Frobel2012}.  With Tat-based export, proteins are not dependent on extracytoplasmic conditions for folding, making the Tat system particularly important for  bacterial virulence \cite{DeBuck2008} and biotechnology applications \cite{Bruser2007}.

Tat translocases \cite{Palmer2012, Frobel2012} consist of the TatA, TatB, and TatC proteins in the bacterial inner membrane. TatBC complexes recognize Tat signal peptides, while TatA multimers associate with  TatBC complexes and are thought to form a transmembrane conduit.  TatA complexes are large, dynamic, and broadly distributed in size \cite{Leake2008, Gohlke2005}. This is thought to allow the Tat system to accommodate a broad-range of folded substrate sizes ($9-142$ kDa substrates with approximately $2$-$7$ nm diameter \cite{Berks2000}). 

Not every substrate with a targeting sequence is translocated.  It is not known precisely what substrate properties determine translocatability, though goodness of fold for natural substrates \cite{DeLisa2003, Matos2008, Panahandeh2008, Maurer2009} or moderate size and hydrophilicity for artificial substrates \cite{Cline2007, Richter2007} appear to be important. The distinction between translocatable and non-translocatable (NT) substrates can also be affected by suppressor mutations of the translocon apparatus \cite{Rocco2012}. How might the Tat translocon avoid blockage due to non-translocatable (NT) proteins and protein complexes? One possibility is that the Tat signal peptide only targets substrates to the translocon if they are well folded --- so-called ``proof-reading'' \cite{Palmer2005}. Such a mechanism might prevent NT substrates from binding to and blocking translocons. However, translocation of artificial substrates with long flexible linkers \cite{Lindenstrauss2009} as well as small unstructured substrates \cite{Richter2007} has been reported.  This implies that targeting of substrates to Tat translocons may not sensitively depend on the nature of the substrate.  Indeed, non-translocatable substrates are observed to associate with the translocon \cite{Panahandeh2008, Richter2005}. Nevertheless,  translocatable \cite{Whitaker2012} and NT \cite{Musser2000, Bageshwar2009} substrates appear to only transiently associate with Tat translocons. This is consistent with the observation that degradation of misfolded Tat substrates appears to be independent of the Tat system \cite{Lindenstrauss2010}. 

The binding affinities or unbinding rates of targeted substrates may depend upon substrate properties, such as whether a particular substrate is well-folded or not.  This is one form of the quality control hypothesis \cite{DeLisa2003}. Lower binding affinities and/or higher unbinding rates would lead to lower translocation of targeted substrates, and presumably higher translocation of the remaining substrates.   Nevertheless, this leaves open the question of whether any such quality control is needed to explain existing experimental phenomenology of the Tat system. 

Accordingly, we explore the quality control null hypothesis --- in which binding and unbinding rates of targeted substrates to the Tat translocon do not  depend on substrate properties.  To do this we develop a stochastic model of the binding and unbinding of protein substrates coupled with TatA cluster dynamics and substrate translocation. We allow a bound substrate to translocate when sufficient TatA are present in a cluster. To allow TatA clusters to dynamically adjust to different substrate sizes, we additionally allow the TatA cluster dynamics to depend on the binding status of the cluster: with a bound substrate, growth of the cluster is favoured, and without a bound substrate, shrinkage is favoured; such behaviour is similar to TatA behaviour seen \emph{in vivo} \cite{Panahandeh2008}.  This also reflects the experimental observation that the TatA oligomerization process is induced by the substrate and this is reversed only once the substrate unbinds or translocates \cite{Alcock2013}.  Consistent with our quality control null hypothesis, we also assume that cluster dynamics do not depend on substrate properties. With our model, both the TatA cluster size distribution and substrate translocation rates are computationally investigated.  The model allows the optimization of the translocation efficiency of multiple Tat translocases as the fraction $f$ of NT substrates is varied.  We find that significant translocation is possible with the non-specific substrate disassociation rate. The model also recovers a notable large-size tail that was observed in high-resolution {\em in situ} fluorescence studies of TatA clusters \cite{Leake2008}, which we ascribe to transiently stalled translocases.  In addition, we develop an analytical three-state model without TatA cluster dynamics. This three-state model provides a reasonable approximation of our full stochastic results, and is consistent with {\em in vitro} translocation data of mixed translocatable and NT substrates \cite{Musser2000}.

\section{Model}
\label{sec:model}

We assume a fixed number $N$ of oligomeric TatA translocation complexes (or ``clusters''), and a fixed total number $n_{tot}$ of TatA molecules in the membrane. At a given time $t$, the $i$-th oligomeric TatA cluster is comprised of $n_i$ monomers. TatA monomers that are not in clusters form a common monomeric pool, where $n_{pool}= n_{tot} - \sum_{i=1}^{N} n_i$.  We note that TatBC clusters (see e.g. \cite{Tarry2009}) are implicit in our model, and are necessary for substrate binding, unbinding, and translocation.

\begin{figure}[t]  
 \begin{center}
\includegraphics[width=3.5in]{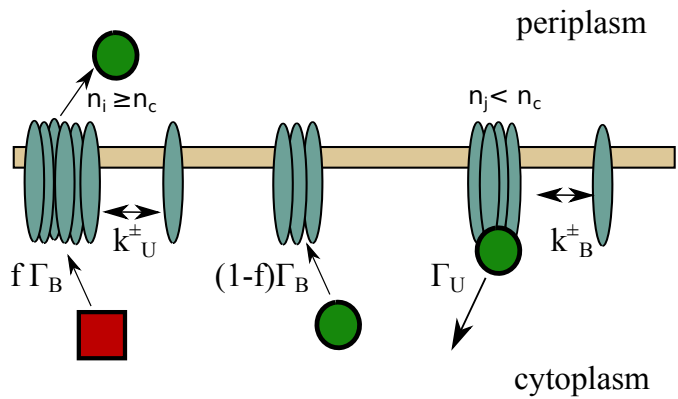}
  \end{center}
\caption{{\bf Illustration of our model dynamics.} Membrane associated TatA (blue-grey ellipses) are in three clusters together with two monomers from $n_{pool}$. Substrates can associate with clusters at rate $\Gamma_B$, and are either translocatable (green circles, fraction $1-f$) or non-translocatable (NT red square, fraction $f$). Translocatable substrates are translocated into the periplasm when the cluster size $n_i$ equals or exceeds a threshold $n_c$. Substrates disassociate from clusters back into the cytoplasm at rate $\Gamma_U$, and this same disassociation rate applies both to translocatable substrates with $n_i<n_c$ and to NT substrates. Clusters grow by a monomer at rate $k^+ n_{pool}$ or shrink at rate $k^-$, and these rates depend on whether the cluster is unassociated (``U'') or bound (``B'') to a substrate.  There are $N$ clusters, and $n_{tot}$ TatA molecules that include both those in clusters and those in $n_{pool}$. TatBC is not shown, but is implicitly part of each cluster to allow substrate association.}
\label{Fig:model}
\end{figure}

The model dynamics are illustrated in Fig.~\ref{Fig:model}. Each TatA cluster may be either associated with a translocation substrate, or not. A substrate has a Tat targeting sequence, or is a complex of molecules associated with a protein with a targeting sequence \cite{Palmer2012}.  Association of substrates to empty clusters occurs at rate $\Gamma_B$, but a fraction $f$ of those substrates are NT. Non-specific unbinding of substrates occurs at a rate $\Gamma_U$, which allows for unblocking of clusters bound to NT substrates but also causes premature release of translocatable substrates. The values of $\Gamma_B$ and $\Gamma_U$ are discussed in Sec.~\ref{subsec:further}. Following our quality control null hypothesis, we assume that the rates $\Gamma_B$ and $\Gamma_U$ do not depend on whether the substrate is translocatable or NT, nor on any non-specific interactions between substrates and the Tat translocon. While the details of the signal peptide can affect translocation rates  \cite{Bageshwar2009, Stanley2000, Hinsley2001}, we assume for simplicity that all of our substrates have the same signal peptides.

Bound translocatable substrates are translocated when their associated cluster is equal to or larger than a critical size $n_c$. For oligomeric substrates \cite{ma10}, $n_c$ would reflect the oligomer size.   NT substrates are those that do not translocate regardless of cluster size, whether due to misfolding, substrate size, or other substrate properties. 

Clusters grow by one monomer at a rate $k^+ n_{pool}$, where $n_{pool}$ is the number of monomers not associated with any cluster, and shrink by one monomer at a rate $k^-$. We assume that there are distinct rates for complexes with bound substrates ($k_B^{\pm}$) and those without ($k_U^{\pm}$). These distinct rates lead to a more dynamic system of growing and shrinking clusters, and allows large substrates to be temporarily accommodated through cluster growth, consistent with TatA recruitment to functionally engaged Tat translocons \cite{Panahandeh2008, Alcock2013}. Because substrates primarily interact with TatB and TatC \cite{Tarry2009, Kostecki2010}, we expect that TatBC modulates TatA cluster growth and so assume that the rates $k^{\pm}$ are independent of cluster size compared to the differences between $k^{\pm}_U$ and $k^{\pm}_B$. Following our quality control null hypothesis, we also assume that the cluster rates $k^{\pm}$ do not vary across substrates.  

Very little is known about effective rate constants for cluster growth.  All of our rates ($k$'s and $\Gamma$'s) are dimensionless; and we generally work in units of the binding rate so that $\Gamma_B=1$.  Unless otherwise indicated, we use $k_B^+=0.05$, $k_B^- = 1.5$, $k_U^+=0.005$, and $k_U^- = 7.0$, where we have $k_B^+ > k_U^+$ and $k_B^- < k_U^-$ so that substrate-associated clusters grow faster and shrink more slowly than unbound clusters. The values for the parameters $k^{\pm}$ are chosen to allow growth and shrinkage of clusters to occur quickly enough to reach $n_c$, but not so rapidly that binding and cluster dynamics are on different timescales. These rates were found to give a cluster size distribution that is qualitatively similar to what is seen experimentally (see below). We systematically vary both the unbinding rate $\Gamma_U$ and the NT fraction $f$. Variation of the parameter values $k^{\pm}$ is explored in the supplemental materials. 

The Gillespie algorithm \cite{Gillespie1977} was used to perform fully stochastic simulations of cluster growth and shrinkage, together with substrate binding, unbinding, and translocation. Clusters were allowed to reach a steady-state distribution before the time-averaged translocation rate $R$ and distribution of cluster sizes $P(n)$ were measured.  Experimental studies of fluorescently labelled TatA indicate that there are $N = 15\pm 9$ TatA clusters per bacterial cell, with approximately $n_{tot} \approx 560$ TatA molecules per cell and $n_{pool} \approx 100$ TatA that are not associated with complexes \cite{Leake2008}. Correspondingly, unless otherwise noted we take the number of clusters $N=15$ and the number of monomers $n_{tot} = 560$. 

\section{Results}
\subsection{Cluster size}
\begin{figure}[t]  
 \begin{center}
\includegraphics[width=3.5in]{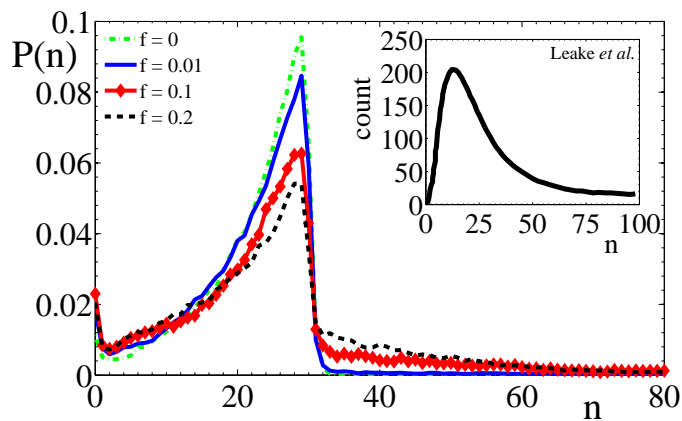}
  \end{center}
\caption{{\bf The cluster size distribution.} The cluster size distribution $P(n)$ vs.\ cluster size $n$ for a variety of NT fractions $f$ with $n_{tot}=560$ and $N=15$. We use $\Gamma_U=0.5$ and a single   threshold size for translocation $n_c=30$. A prominent peak is seen at $n_c$, since unassociated clusters tend to shrink after translocation because $k^-_U>k^-_B$ and $k^+_U < k^+_B$ . As $f$ increases, a significant tail grows above $n_c$ due to growth of clusters with associated NT substrates. The inset shows the experimental distribution of TatA clusters digitized from Fig.~4I of Leake {\em et al} \protect\cite{Leake2008}. The parameter values of this figure, together with $\Gamma_B=1$, $\Gamma_U=0.5$, $k_B^+=0.05$, $k_B^- = 1.5$, $k_U^+=0.005$, and $k_U^- = 7.0$, also apply to other figures unless otherwise noted.}
\label{Fig:dist}
\end{figure}

The model cluster size distributions plotted in Fig.~\ref{Fig:dist}, of $P(n)$ vs.\ the cluster size $n$, exhibit some of the qualitative features seen in the experimental distribution of TatA cluster sizes reported by Leake {\em et al} \cite{Leake2008} and shown in the inset: an increase from arbitrarily small clusters, a distinct peak, and an extended tail for larger cluster sizes. 

As shown in Fig.~\ref{Fig:dist}, increasing the NT fraction $f$ increases the magnitude of the tail of $P(n)$ for $n>n_c$.  Clusters with bound substrates will tend to grow, due to larger $k^+_B$ and smaller $k^-_B$. Clusters with bound NT substrates do not translocate at $n_c$, so growth of individual clusters beyond $n_c$ will be limited by the unbinding rate $\Gamma_U$ (see Fig. S1). Additionally, depletion of the TatA monomer pool generally limits cluster growth. Systematically larger tails for $P(n)$ are seen with more rapid bound growth $k_B^+$ (see Fig.~S2), and with smaller unbound decay $k_U^-$ (see Fig.~S3).  Interestingly, $P(n)$ is also affected by changes in $\Gamma_B$ (see Fig.~S1). Larger $\Gamma_B$ allows rapid rebinding of substrates to large clusters, which prevents their relaxation and leads to a larger tail for $n>n_c$ and less weight for $n<n_c$.  

\begin{figure}[t] 
 \begin{center}
\includegraphics[width=3.5in]{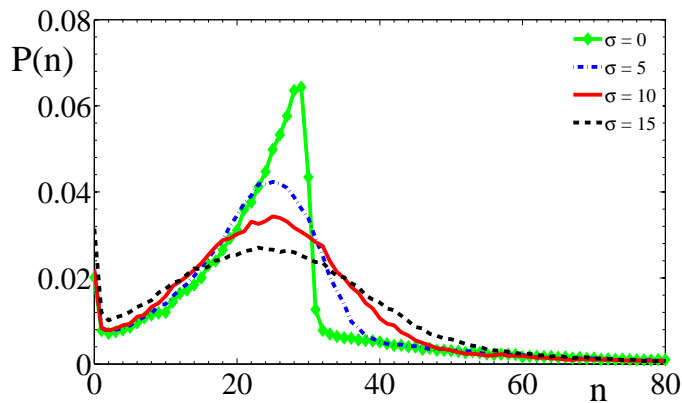}
  \end{center}
\caption{{\bf Cluster size distribution with a distribution of substrate sizes.} $P(n)$ vs.\ cluster size $n$ for different distributions of the substrate sizes $n_c$. All substrate size distributions are Gaussian distributions with mean $\langle n_c \rangle =30$, truncated at $2 \sigma$. In all cases $f=0.1$, so $\sigma=0$ (green diamonds) is the same as the red diamonds in Fig.~\protect\ref{Fig:dist}. Other parameter values are the same as in Fig.~2. }
\label{Fig:distgauss}
\end{figure}

In addition to the tail of $P(n)$ for $n>n_c$, the location of the peak of $P(n)$ in the model is approximately determined by $n_c$.  A variety of Tat substrates \cite{Palmer2012}, with a range of sizes and abundances, would be expected to round the sharp peak obtained with the model using a single value of $n_c$ and lead to better qualitative agreement with the rounded experimental distribution of TatA cluster sizes \cite{Leake2008} shown in the inset of Fig.~\ref{Fig:dist}. To explore this effect in Fig.~\ref{Fig:distgauss}, for each substrate we have selected $n_c$ from a truncated Gaussian distribution with standard deviation $\sigma$, truncated at $2 \sigma$ with an average $\langle n_c \rangle=30$. We see that using a distribution of $n_c$ rounds the peak of the cluster size distribution $P(n)$ but does not significantly change the behaviour for $n<n_c$.   Since we are using a single value of $n_c$ for the rest of this paper, we have chosen $n_c=30$ to better emphasize the $n>n_c$ tail of $P(n)$.  

\begin{figure}[t] 
 \begin{center}
 \begin{tabular}{c}
\includegraphics[width=3.5in]{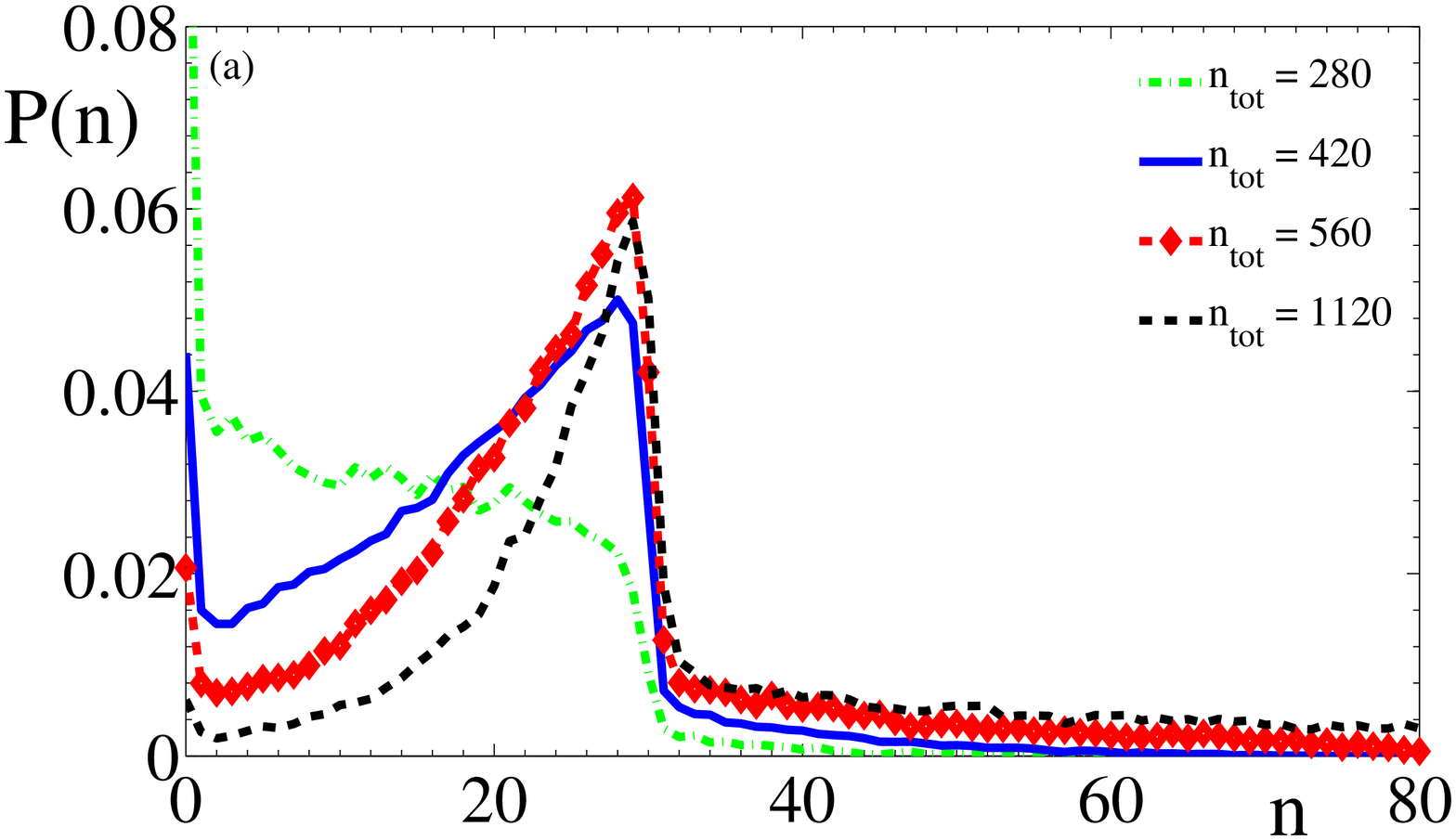}
\\
\\
\includegraphics[width=3.5in]{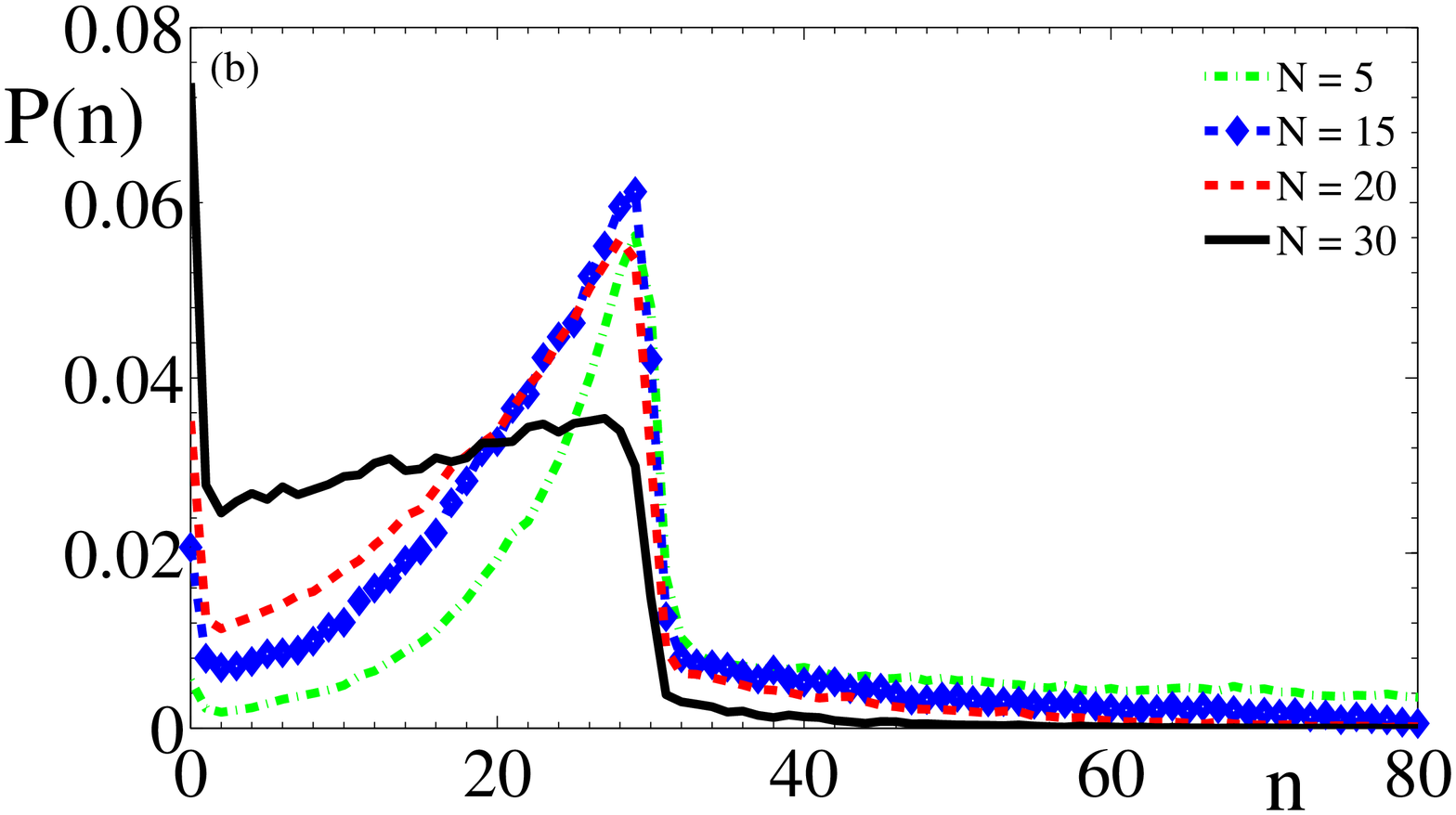}
\end{tabular}
  \end{center}
\caption{{\bf Cluster size distribution as number of clusters or number of TatA are varied.} (a) Cluster size distribution $P(n)$ vs.\ $n$ for $N=15$ clusters with variable $n_{tot}$. (b) as previous, but with $n_{tot}=560$ and variable $N$. For both figures we use $f=0.1$. Other parameter values are as given in Fig.~2.}
\label{Fig:P(n)vsntot} 
\end{figure}

 In Fig.~\ref{Fig:P(n)vsntot} we show how the cluster size distribution $P(n)$ changes as (a) the number of TatA monomers $n_{tot}$ or (b) the number of translocons $N$ is changed.  For the smallest $n_{tot}=280$ the peak near $n_c$ is lost. For larger $n_{tot} \gtrsim 420$ the distribution at small $n$ retains a similar shape with increasing $n_{tot}$, but decreases in magnitude as $n_{tot}$ increases. In contrast, the large-$n$ tail increases with $n_{tot}$.  As shown in Fig.~\ref{Fig:P(n)vsntot}(b), we see corresponding effects when $N$ is varied at fixed $n_{tot}$, with similar monomer numbers per translocon, $n_{tot}/N$, leading to qualitatively similar cluster size distributions. 

\subsection{Translocation rate}
\begin{figure}[t] 
 \begin{center}
\includegraphics[width=3.5in]{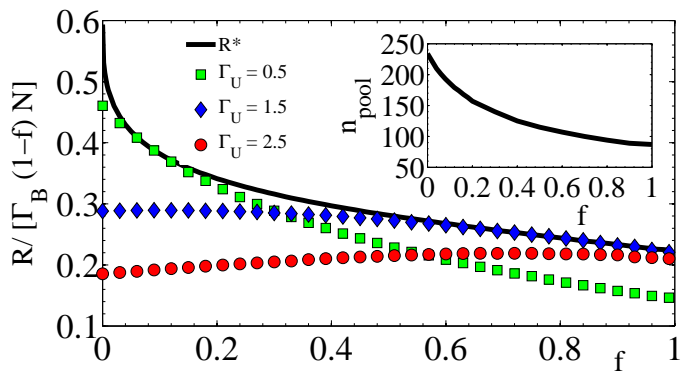}
  \end{center}
\caption{{\bf Scaled translocation rate as NT fraction is varied.} Scaled translocation rate per cluster $R/( \Gamma_B (1-f) N)$ vs.\ the NT fraction $f$ for  $n_{tot}=560$ and $N=15$.  Shown are different unbinding rates $\Gamma_U$.  Other parameter values are as given in Fig.~2. The thick black line is the scaled optimal translocation rate $R^\ast$, where for each value of $f$ we have chosen the $\Gamma_U=\Gamma_U^\ast(f)$ that maximizes $R$.  The inset shows the average size of the monomeric pool, $n_{pool}$ vs.\ $f$, corresponding to the maximal $R^\ast$ line.}
\label{Fig:R}
\end{figure}

The total translocation rate $R$ can be no more than $N \Gamma_B (1-f)$ --- the binding rate of translocatable substrates to $N$ empty clusters. In Fig.~\ref{Fig:R}, we show the scaled translocation rate per cluster, $R/( \Gamma_B (1-f) N) $, vs.\ the NT fraction $f$ for  $n_{tot}=560$ and $N=15$.  As expected, scaled translocation rates are  below the theoretical limit of $1$ for all NT fractions $f$. This is because, as illustrated in Fig.~\ref{Fig:dist}, most clusters are smaller than $n_c$ --- and so need to grow before they can translocate bound substrates.  Translocation is further reduced by bound NT substrates, which must be disassociated before further substrate binding and translocation is possible. We show three scaled translocation rates, for $\Gamma_U= 0.5$, $1.5$, and $2.5$ (green squares, blue diamonds, and red circles, respectively). We see that despite no specific quality control mechanism, translocation rates of $50\%$  of the theoretical maximum are possible at smaller $f$ while even at large $f$ translocation rates can still reach $20\%$ of the theoretical maximum. 

\begin{figure}[t]
 \begin{center}
 \begin{tabular}{c}
\includegraphics[width=3.5in]{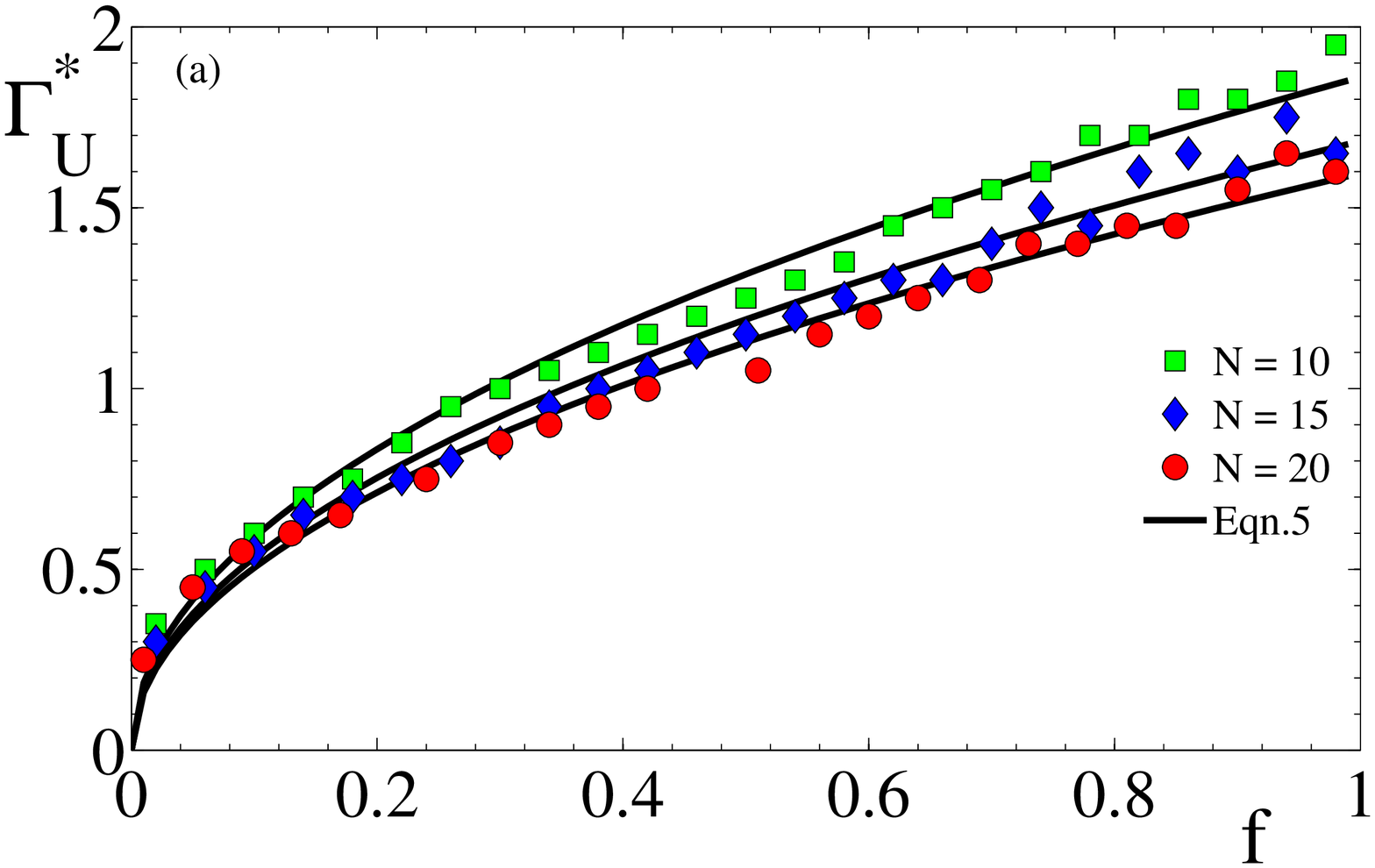} \\
\\
\includegraphics[width=3.5in]{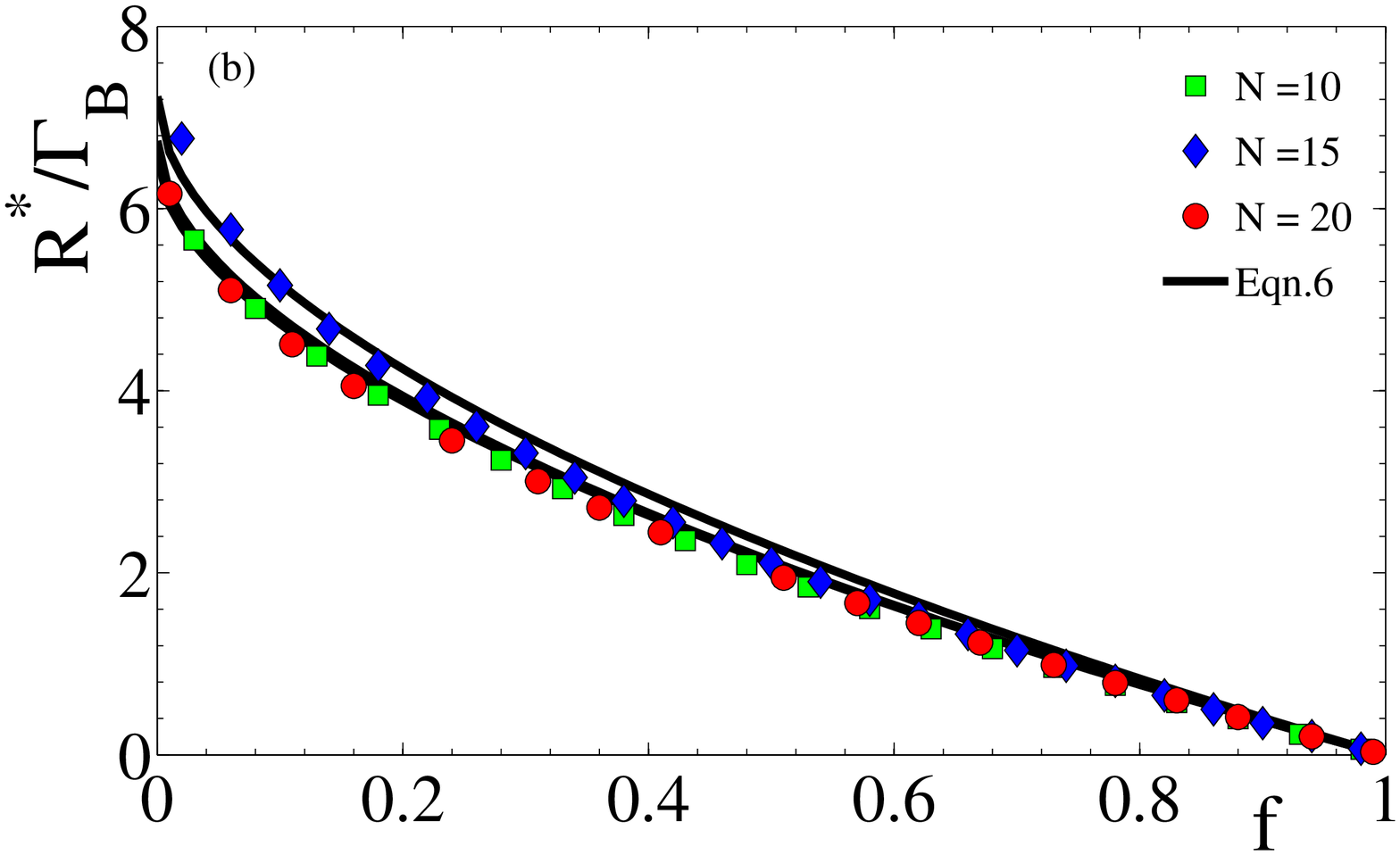}
\end{tabular}
  \end{center}
\caption{{\bf Optimized translocation as NT fraction is varied.} (a) The optimal unbinding rate, $\Gamma_U^\ast$, that maximizes the translocation rate vs.\ the NT fraction $f$. For a fixed number of TatA, $n_{tot}=560$, results are shown for different numbers of clusters $N$ as indicated by the legend.  Other parameter values are as given in Fig.~2. All curves show a characteristic square-root dependence on $f$, as shown by the phenomenological fits to a simplified three-state model of translocation from Eqn.~\ref{EQN:Gammuast} (black lines). (b) The corresponding scaled total translocation rates $R^\ast/\Gamma_B $ vs.\ $f$, together with phenomenological fits (black lines) to  Eqn.~\ref{EQN:Rast}.}
\label{Fig:GammaU}
\end{figure}

From the change in the translocation rate dependence on $f$ between the different $\Gamma_U$ values in Fig.~\ref{Fig:R}, we see that for smaller values of $f$ a smaller $\Gamma_U$ leads to higher scaled translocation rates while at larger values of $f$ a larger $\Gamma_U$ leads to more translocation. More generally, we find that there is an optimal $\Gamma_U^\ast$ that leads to the highest translocation rate $R^\ast$ for each value of the NT fraction $f$. If $\Gamma_U < \Gamma_U^\ast$, then too many NT substrates block translocation. If $\Gamma_U> \Gamma_U^\ast$, then too many translocatable substrates are removed before the cluster size reaches $n_c$. We numerically identify $\Gamma_U^\ast$  by varying $\Gamma_U$ and measuring $R$, as $f$ is varied. The corresponding optimal translocation rate $R^\ast$ is plotted with a solid black line in Fig.~\ref{Fig:R}. The curves for specific values of $\Gamma_U$ in Fig.~\ref{Fig:R} demonstrate that close to optimal translocation can be obtained for a wide range of $f$ for each value of $\Gamma_U$.  The relationship between $\Gamma_U^\ast$ and $f$ is shown in Fig.~\ref{Fig:GammaU}(a).

We also note that there is a decreasing monomeric pool of TatA with increasing $f$, as shown in the inset of Fig.~\ref{Fig:R} and reflecting the increased tail of $P(n)$ with $f$ shown in Fig.~\ref{Fig:dist}. Our observed range of $n_{pool} \sim 100-200$ is comparable to $n_{pool} \approx 100$ reported experimentally \cite{Leake2008}.

\subsection{Optimal number of clusters $N$}
\begin{figure}[t] 
 \begin{center}
 \begin{tabular}{c}
\includegraphics[width=3.5in]{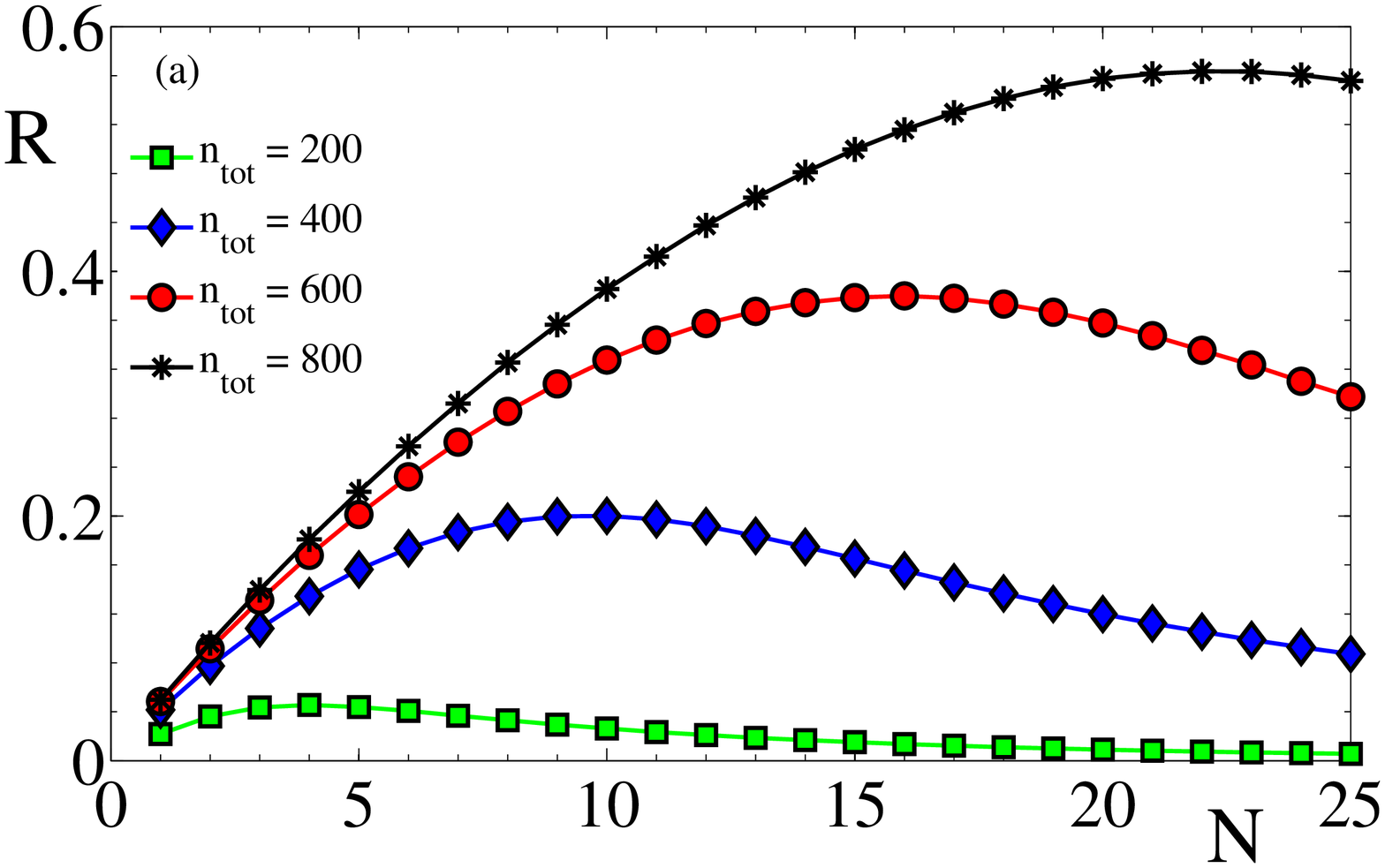} \\
\\
\includegraphics[width=3.5in]{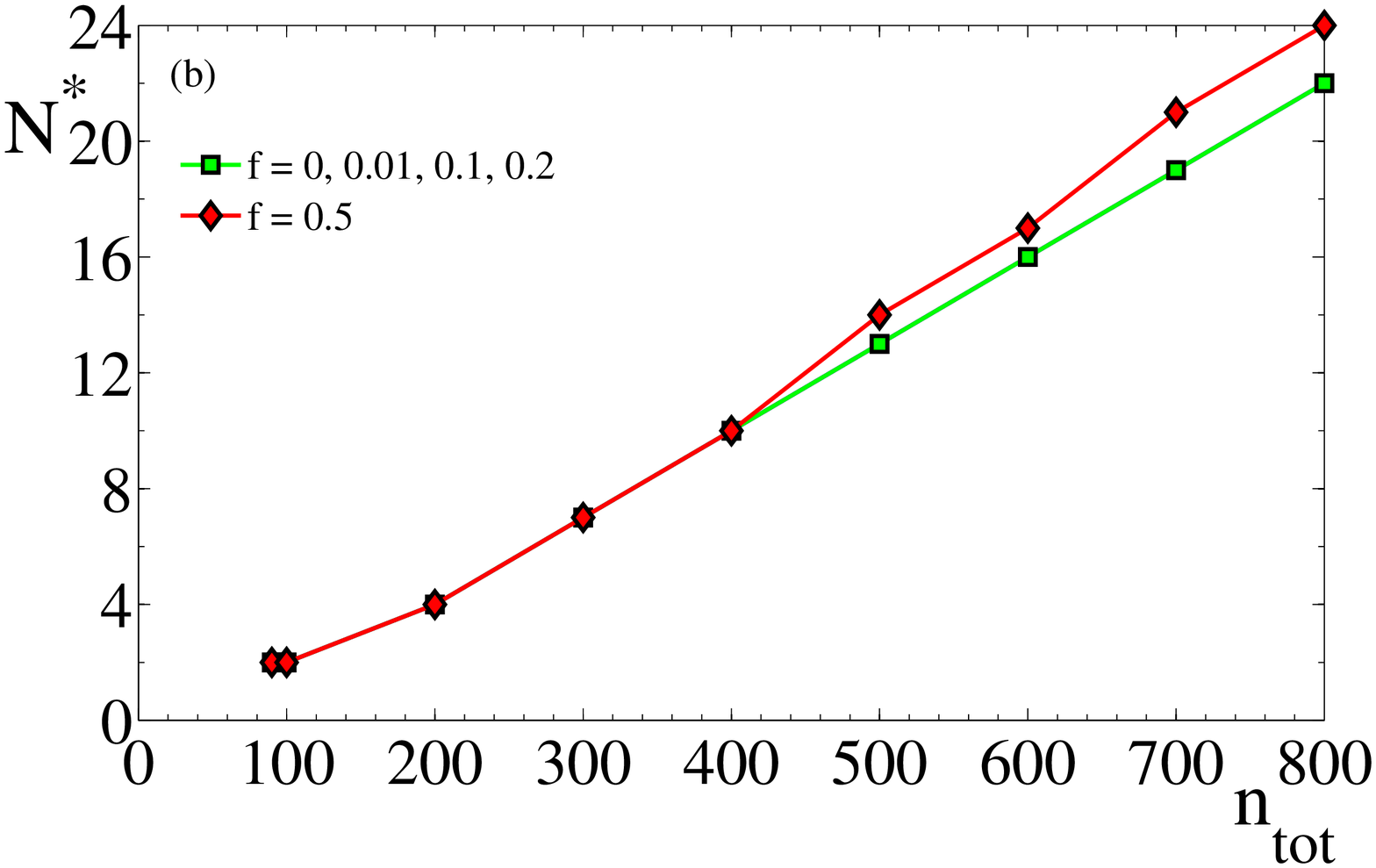}
\end{tabular}
  \end{center}
\caption{{\bf Translocation as the number of clusters is varied.} (a) The total translocation rate, $R$, vs.\ number of clusters $N$ for variable total number of monomers $n_{tot}$.  The NT fraction $f=0.1$. (b) The number of clusters that maximizes translocation, $N^\ast$, versus $n_{tot}$ for different NT fractions $f$. The green squares have four superimposed curves. Other parameter values are as given in Fig.~2.}
\label{Fig:RvsN} 
\end{figure}

In Fig.~\ref{Fig:GammaU}(a), the optimal $\Gamma_U$ vs $f$ is investigated for different number of clusters $N$. The optimal $\Gamma_U^\ast$ depends strongly on $f$ but only weakly on $N$, and a similar weak dependence is seen in  Fig.~\ref{Fig:GammaU}(b) for the corresponding optimal scaled  total translocation rates $R^\ast/\Gamma_B$. Nevertheless, we can see that $R^\ast$ does not monotonically increase with $N$, but is slightly larger for $N=15$ (blue diamonds). This is explored in more detail in Fig.~\ref{Fig:RvsN}(a), where the total translocation rate $R$ is plotted vs.\ number of clusters $N$ for different total number of TatA, $n_{tot}$, as indicated by the legend. For each $n_{tot}$ there is an optimal $N^\ast$ that maximizes the total translocation rate $R$. This behaviour arises because the clusters share the same fixed number  $n_{tot}$ of TatA, forcing a tradeoff between cluster size and cluster number.   A smaller $N$ has a reduced total translocation rate, because $n_{pool}$ is sufficiently large for translocation to be limited by $\Gamma_B$, while a larger $N$ will slow the cluster growth needed to translocate at size $n_c$ due to the depletion of $n_{pool}$. 
  
Fig.~\ref{Fig:RvsN}(b) shows that the optimal $N^\ast$ increases with $n_{tot}$. For $n_{tot} \gtrsim 200$, the optimal number of clusters $N^\ast$ scales approximately linearly with larger values of $n_{tot}$.  We also see that $N^\ast$ is not strongly dependent on the NT fraction. Together, this implies that stoichiometric control of cluster number may be sufficient to maintain close to optimal translocation rates in the face of varying levels of TatA. 

We can now reconsider how the cluster size distribution $P(n)$ depends upon $N$, as shown in Fig.~\ref{Fig:P(n)vsntot}(b). With the chosen parameter values the optimal number of clusters is $N^\ast = 15$.  For  $N \lesssim N^\ast$, increasing $N$ slightly decreases the tail at larger $n$ due to the availability of monomers,  and moderately increases the distribution of smaller $n$.  However, once $N > N^\ast$ we see a qualitatively different distribution with a significant number of very small clusters.  In comparison with the experimental $P(n)$ (see inset of Fig.~2) we see that our model gives similar results if $N \lesssim N^\ast$ --- both have few small clusters and an increase to a distinct peak, rather than a larger population of smaller clusters.  The model parameters are underconstrained by the current experimental data, and we have chosen kinetic rates so that $N^\ast=15$ is similar to the number of TatA clusters observed {\em in vivo} \cite{Leake2008} and equal to our default value of $N$. Nevertheless, we can say that results are consistent with the Tat system having close to an optimal number of translocons given the amount of TatA.

\subsection{Three-state model}
\label{subsec:threestate}
To better understand our translocation rates, we consider a simplified stochastic three-state Tat translocation model.  The three states, with corresponding probabilities, are empty ($p_0$), bound with translocatable substrate ($p_B$), and bound with non-translocatable substrate ($p_{NT}=1-p_0-p_B$). Two transition rates that directly correspond to the full model are $\Gamma_B$, the rate at which substrates bind to an empty cluster, and $\Gamma_U$, the rate at which substrates unbind from a cluster without translocation. We add one additional rate, $\Gamma_T$, the translocation rate of translocatable substrates, to phenomenologically account for cluster size dynamics, threshold size $n_c$, as well as the mechanics of translocation.  Our dynamical equations are then
 \begin{eqnarray}
 		\frac{dp_0}{dt} &=& \Gamma_U p_{NT} + (\Gamma_U+\Gamma_T) p_B-\Gamma_B p_0, \\
	\frac{dp_B}{dt} &=& (1-f) \Gamma_B p_0-(\Gamma_U+\Gamma_T) p_B, \\
	\frac{dp_{NT}}{dt} &=& f \Gamma_B p_0- \Gamma_U p_{NT},
\end{eqnarray}
where $f$ is the fraction of non-translocatable substrates. We use $p_0+p_B+p_{NT}=1$ to solve these equations in steady state, where the time-derivatives vanish, and obtain
\begin{equation}
	p_B=\frac{(1-f) \Gamma_B \Gamma_U}{f \Gamma_B \Gamma_T+\Gamma_U(\Gamma_B+\Gamma_T+\Gamma_U)},
\end{equation}
and a corresponding translocation rate $R = N_{eff} \Gamma_T p_B$, where $N_{eff}$ is the effective number of independent clusters participating in translocation. Note that our three-state model mathematically corresponds to a special case of a `one site' model of an always-open transport channel developed by Zilman {\em et al} \cite{Zilman2010}, where their $J_n= (1-f) \Gamma_B$, $J_m = f \Gamma_B$, $r_m^{right}=0$, $r_m^{left} = r_n^{left}=\Gamma_U$, and $r_n^{right}=\Gamma_T$. 

To compare with optimized translocation results in Fig.~\ref{Fig:GammaU}, we maximize the translocation rate with respect to the unbinding rate $\Gamma_U$, with $\partial R/\partial \Gamma_U=0$, and obtain
\begin{eqnarray}
	\Gamma_U^\ast &=& \sqrt{\Gamma_T \Gamma_B f},\label{EQN:Gammuast} \\
	R^\ast &=& \frac{N_{eff} \Gamma_T \Gamma_B (1-f)}{\left[\Gamma_B(1-f) + (\sqrt{\Gamma_T} + \sqrt{\Gamma_B f})^2 \right]}. \label{EQN:Rast}
\end{eqnarray}
$N_{eff}$ is the effective number of independent clusters participating in translocation, so that we expect $N_{eff} \lesssim N$.  As shown by the solid black lines in Fig.~\ref{Fig:GammaU}(a), the characteristic square-root dependence of $\Gamma_U^\ast$ vs.\ $f$ from Eqn.~\ref{EQN:Rast} describes the full model results well. Fit by eye, we find that $\Gamma_T = 3.5, 2.8$, and $2.5$ for $N=10, 15$, and $20$ respectively. The phenomenological translocation rate for each cluster, $\Gamma_T$, decreases with increasing $N$ --- as expected since $n_{pool}$ decreases with increasing $N$. As shown by the solid black lines in Fig.~\ref{Fig:GammaU}(b), Eqn.~\ref{EQN:Rast} also provides a satisfactory fit for the optimal translocation rates. Using the  $\Gamma_T$ values, we fit by eye to find $N_{eff} = 6.2, 7.0$, and $6.4$, for $N=10, 15$, and $20$, respectively. $N_{eff} < N$, as expected. Interestingly, we see that $N_{eff}$ is largest for $N=15$, where $N \simeq N^\ast$.  Note that the values of $\Gamma_T$ and $N_{eff}$  will depend on the kinetic parameters ($k^{\pm}$) of the full model.

Our three-state model does not include any cluster size dynamics, so is largely independent of our detailed assumptions of cluster size dynamics. Additionally, it appears to be a reasonable approximation of our full dynamical model.  The three-state model is also easily adapted to different experimental protocols.

Our models parameterize the non-translocatable fraction $f$ independently from the substrate binding rate $\Gamma_B$; this highlights the role of $f$ in determining an optimal $\Gamma_U$. \emph{In vitro}, it can be more convenient to independently adjust the bulk substrate concentrations, $\rho_T$ and $\rho_{NT}$ of translocatable and NT substrates, respectively. For our parameters, this then gives
\begin{eqnarray}
	\Gamma_B &=& \gamma_T \rho_T +  \gamma_{NT} \rho_{NT} \\
	f &= &\frac{1}{1+ \rho_T/ \rho_{NT}}.
\end{eqnarray}
We have allowed for distinct binding constants $\gamma_T$ and $\gamma_{NT}$ for translocatable and NT substrates, respectively, to allow for quantitative effects of substrate size in affecting diffusion-limited association rates \cite{Berg1985}. We will continue to assume that unbinding $\Gamma_U$ is the same for both substrates.

Musser and Theg quantitatively characterized translocation in a thylakoid Tat system \cite{Musser2000}. They considered radioactively labelled ($\rho_R =100$ nM) and unlabelled ($\rho_{NR}$) fractions of a translocatable substrate, so that $\rho_T = \rho_R +\rho_{NR}$ and $\rho_{NT}=0$. Then the total translocation amount of labelled substrate in time $\Delta t$ is $R_{tot,R} = R \Delta t \rho_R/\rho_T$, i.e.
\begin{equation}
	R_{tot,R}= \frac{ \gamma_T \rho_R \Gamma_T N_{eff} \Delta t}{\gamma_T ( \rho_R +\rho_{NR})+\Gamma_U+\Gamma_T},
	\label{EQN:RR}
\end{equation}
where we have used our unoptimized three-state model with a phenomenological translocation rate $\Gamma_T$ and $f=0$.  

Musser and Theg then repeated their experiment with a non-translocatable (biotinylated) unlabelled substrate concentration $\rho_{NT}=\rho_{NR}$, and the same translocatable labelled concentration so that $\rho_T=\rho_R$. We then have a total translocation amount of labelled substrate
\begin{equation}
	R_{tot,T} = \frac{\gamma_T \rho_R \Gamma_T N_{eff} \Delta t}{\gamma_T  \rho_R+\gamma_{NT} \rho_{NR}+ \Gamma_U+\Gamma_T + 	\gamma_{NT} \rho_{NR} \Gamma_T/\Gamma_U},
	\label{EQN:RT}
\end{equation}
with an additional term in the denominator due to $\gamma_{NT}$. We see that $R_{tot}^{-1}$ in both Eqn.~\ref{EQN:RR} and \ref{EQN:RT}  depends linearly on the unlabelled concentration $\rho_{NR}$.

\begin{figure}[t] 
 \begin{center}
	\includegraphics[width=3.5in]{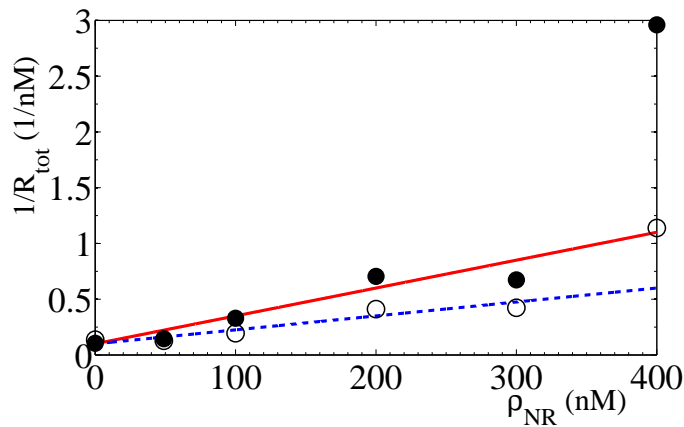}
  \end{center}
\caption{{\bf \emph{In vitro} translocation data analyzed with simplified three-state model.} The inverse total translocation vs.\ unlabelled biotinylated substrate concentration $\rho_{NR}$ from Fig.~3E of Musser and Theg \protect\cite{Musser2000}. All points have $\rho_R=100$nM. The open and closed circles correspond to experiments done with avidin (where $\rho_{NT}=\rho_{NR}$, and Eqn.~\ref{EQN:RT} is applicable) or without (where $\rho_{NT}=0$, so that Eqn.~\ref{EQN:RR} is applicable), respectively. We have shown linear fits by eye to Eqns.~\ref{EQN:RR} or \ref{EQN:RT} with solid red or dashed blue lines, respectively. We have imposed a common y-intercept on our fits, corresponding to the translocation of radioactively labelled substrate $\rho_R$ that remains  the same between the two experiments. }
\label{Fig:Musser} 
\end{figure}

In Fig.~\ref{Fig:Musser}, we have plotted the inverse total translocation amount $R_{tot}^{-1}$ vs.\ the concentration of unlabelled substrate $\rho_{NR}$ using  digitized experimental data from Fig.~3E of \cite{Musser2000}.   The expected linear behavior of $R_{tot}^{-1}$ vs $\rho_{NR}$ is apparent at larger values of $R_{tot}$ (i.e. smaller values of $\rho_{NR}$) , where systematic and statistical errors should be less significant.  In that regime, our three-state model appears consistent with the experimental translocation data both with and without NT substrates.

\section{Discussion and conclusions}
\label{subsec:further}

The Tat protein export pathway translocates folded proteins across membranes in bacteria and plant cells \cite{Palmer2012,Frobel2012}, and Tat translocons must accommodate a range of substrate sizes while staying unclogged by NT substrates. Our model demonstrates  how a non-specific substrate unbinding rate ($\Gamma_U$) can recover an appreciable fraction of the maximal translocation rate achievable with no NT substrates (see Fig.~\ref{Fig:R}). Non-specific substrate unbinding still kinetically discriminates \cite{Howan2012} between translocatable and NT substrates, since the former are often translocated before unbinding can occur.

Association and disassociation rates of one substrate with the Tat translocation complex have recently been estimated {\em in vitro} by Whitaker {\em et al} \cite{Whitaker2012} with bacterial extracts. They found $\Gamma_U \approx 0.042 $s$^{-1}$ (their $k_{off}$) and $\Gamma_B = k_{on} \rho_{substrate}$, where $k_{on} \approx 10^6 - 10^7 M^{-1}$s$^{-1}$ \cite{Whitaker2012}. (This binding may occur via a membrane-associated intermediate \cite{Bageshwar2009}.) Given a bacterial volume of $1 \mu$m$^{3}$, substrate numbers of $1-10^4$ per cell would give $\Gamma_B \approx 10^{-3}/s - 10^2/$s per distinct substrate.  Given the number of different Tat substrates in the cell \cite{Palmer2012}, it appears that $\Gamma_U \lesssim \Gamma_B$ is typical for, e.g., {\em Escherichia coli}, but $\Gamma_U \gtrsim \Gamma_B$ may be accessible in, e.g., model vesicular systems with few translocation substrates. We see from Figure \ref{Fig:GammaU}(a) that $\Gamma_U\lesssim\Gamma_B$ (where $\Gamma_B=1$), is consistent with optimized translocation for NT fractions $f \lesssim 0.2$.

We have also explored optimization of the number of clusters $N$ as the number of TatA molecules in the membrane $n_{tot}$ is varied.   The number of clusters has an optimal value, $N^\ast$, which maximizes the translocation rate corresponding to a given number of TatA molecules. For $n_{tot}=560$ monomers, $N^\ast \approx 15$, which leaves a pool of monomers $n_{pool} \approx 100$. These optimal values are consistent with {\em in vivo} studies of the Tat system \cite{Leake2008}.  We also find that if the number of clusters is much more than $N^\ast$ then the distinctive peak of the cluster size distribution seen in Fig.~\ref{Fig:dist}, and reported experimentally by Leake {\em et al} \cite{Leake2008}, is lost (see solid black curve in Fig.~\ref{Fig:P(n)vsntot}(b)). Together this indicates that $N$ could be close to optimal in the bacterial Tat system. It would be interesting  to explore how, and how well, optimal behaviour is achieved {\em in vivo} as $f$ and $n_{tot}$ are varied. 

Our full model, with cluster size dynamics that depend on whether substrates are bound or not, recovers the qualitative shape and distinctive long tail of large TatA clusters reported by Leake {\em et al} \cite{Leake2008}.  Our model is a quantitative ``bespoke channel model'' \cite{Palmer2012}, in which the cluster size $n_i$ dynamically accommodates  the substrate size $n_c$ so that translocation can occur. We note however that both $n_i > n_c$ and $n_i <n_c$ are observed in our model, the former as a result of recent interactions of the cluster with larger or NT substrates, and the latter as a result of assembly after recent binding and disassembly after unbinding.   We also observe significant variability of our cluster distribution $P(n)$ with number of clusters $N$ and with number of monomers $n_{tot}$. This indicates caution must be taken in interpreting how cluster sizes observed through, e.g., cross-linking studies \cite{DabneySmith2006} or fluorescence microscopy \cite{Leake2008}, respond to substrate sizes. We also note that substrate shape can affect both the critical cluster size $n_c$ and the effective translocation rate $\Gamma_T$ \cite{Whitaker2013}. We find that the cluster size distribution is broadened by a distribution of the critical size $n_c$ (see Fig.~\ref{Fig:distgauss}). The cluster size $n_c$ necessary for translocation may also have a non-linear dependence on substrate radius. However, we do find that the peak of $P(n)$ is always near $n_c$  --- so studies of Tat translocation in a thylakoid system that allows for a single translocation substrate (see e.g. \cite{Musser2000} or \cite{DabneySmith2006}) but with fluorescently labelled TatA (see e.g. \cite{Leake2008}) that resolves the full cluster size distribution should be able to determine how TatA cluster sizes respond to substrate size.

There are many dynamical processes that could couple TatA complex size with substrate translocation; we have implemented a relatively simple one with $k^{\pm}$.  The current picture appears to be that TatBC complexes, with perhaps some TatA, associate with substrates and then recruit more TatA.  Substrates appear to associate at the side of TatBC complexes \cite{Tarry2009}, which could allow for discrimination between bound $k^\pm_B$ and unbound $k^\pm_U$ rates \cite{Alcock2013}.  TatA association with substrate-associated TatBC complexes may happen through recruiting TatA tetramers \cite{Leake2008},  perhaps after initial recruitment of larger TatA complexes (see, e.g., \cite{Muller2005}). Smaller monomer or tetramer association after initial complex recruitment would be consistent with our model, though association of larger TatA complexes at later stages would probably change our cluster size distribution significantly.  Nevertheless, our simplified three-state translocation model, with substrate unbinding but without cluster dynamics, appears to fit translocation data with a variable fraction of labelled substrates \cite{Musser2000}. We believe that as long as both translocation and cluster size dynamics are fast compared to $\Gamma_U$, the translocon can kinetically discriminate \cite{Howan2012} between translocatable and non-translocatable substrates. 

Absolute translocation rates have been measured {\em in vitro} with plant thylakoids, and Alder and Theg \cite{Alder2003} report $v_{max}=6.2/({\rm thylakoid} \cdot {\rm s})$ and $K_m=189{\rm nM}$. With approximately $15000$ translocons per chloroplast \cite{Asai1999} and approximately $500$ thylakoids per chloroplast \cite{Antal2013}, we estimate $N_{eff} \simeq 30$. Comparing with Eqn.~9 with $f=0$ we obtain $v_{max}= \Gamma_T N_{eff}$ and $K_M = (\Gamma_U+\Gamma_T)/\gamma$, which then gives $\Gamma_T \simeq 0.2/$s (assuming $\Gamma_U \ll \Gamma_T$) and $\gamma = 10^6$ M$^{-1}$ s$^{-1}$. This $\gamma$ is in remarkable agreement with the results of Whitaker {\em et al} \cite{Whitaker2012}, and indicates that Tat  kinetics may be similar in the thylakoid and bacterial systems. While this $\Gamma_T \simeq 0.2/s$ is much faster than $\Gamma_U \approx 0.042/s$ \cite{Whitaker2012}, it  somewhat less than the $\Gamma_B$ expected for the more abundant bacterial Tat substrates. This indicates that the most abundant substrates may be rate limited by translocation timing, which includes TatA cluster dynamics, rather than by association or stalled translocons due to NT substrates. 

We do not speculate about the mechanics of a substrate actually crossing the membrane or how a threshold number of TatA would allow translocation for a  substrate of a given size. With respect to translocation, the simple assumptions made are that translocation can occur for a translocatable substrate once a sufficient number of TatA monomers have accumulated, and that the translocation process is rapid compared to the timescales of TatA and substrate binding and unbinding. We focus on the role of an unbinding rate to avoid clogging the Tat translocon and how non-translocatable substrates could affect the translocation of other substrates through the availability of TatA monomers.

We have mostly considered both binding $\Gamma_B$ and unbinding $\Gamma_U$ to be independent of the substrate, and in particular of whether the substrate is translocatable or NT.  But, experimental studies have shown that variations of the signal peptides can affect translocation rates \cite{Bageshwar2009, Stanley2000, Hinsley2001} and that folding or lack of folding can affect signal peptide binding \cite{Panahandeh2008}. Within our model this could arise from variations of $\Gamma_B$, or of $\Gamma_U$, or of both.  Indeed, to best agree with \emph{in vitro} translocation studies of Musser and Theg \cite{Musser2000} in Sec.~\ref{subsec:threestate} we have allowed $\Gamma_B$ to reflect different size substrates. Clearly some substrate dependence of at least $\Gamma_B$ is indicated, and diffusion-limited rates will depend somewhat on the folded nature of the substrate. Substrate dependence of $\Gamma_B$ and $\Gamma_U$, or of cluster-size dynamics through $k^{\pm}$, are straight-forward to include in our model.  Nevertheless, little is known about any such substrate-dependent kinetics, and so we have not attempted to characterize them in our model.

If and how quality control \cite{DeLisa2003} of Tat translocation is achieved is being actively investigated: chaperones can affect Tat substrate binding  \cite{Jack2004}; unstructured proteins \cite{Richter2007} and small, unfolded, hydrophilic polypeptides \cite{Robinson2011} can be translocated by Tat; and mutations in the Tat system can enable it to translocate previously untranslocatable proteins \cite{Rocco2012}. These studies focus on what defines non-translocatable (NT) substrates, while we focus on how reversible substrate association (via $\Gamma_U$) can lead to a significant fraction of the maximal translocation rate being achieved -- see Fig.~\ref{Fig:R}.  We have shown that the optimal unbinding rate depends on the NT fraction $f$. \emph{In vivo}, with a variety of substrates, each with its own abundance and NT fraction, we expect that translocation could be further controlled with substrate-dependent association and unbinding rates. This would amount to specific but passive quality control, and might be implemented in part through variations of associated signal peptides \cite{Bageshwar2009, Stanley2000, Hinsley2001}.  A functional definition of ``non-translocatable'' may also depend upon the speed of translocation with respect to the unbinding rate. It would be interesting to measure substrate unbinding rates in suppressor mutants of Tat translocation that allow previously NT substrates to significantly translocate \cite{Rocco2012}.

\section{Acknowledgments}
We thank the Natural Science and Engineering Research Council (NSERC) for operating grant support, and the Atlantic Computational Excellence Network (ACEnet) for computational resources. CRN thanks ACEnet for fellowship support. AIB thanks NSERC, ACEnet, the Sumner Foundation, and the Killam Trusts for fellowship support.

\section*{References}
\bibliographystyle{unsrt}
\bibliography{references}

\begin{thebibliography}{10}

\bibitem{Economou2006}
A~Economou, P~J Christie, R~C Fernandez, T~Palmer, G~V Plano, and A~P Pugsley.
\newblock Secretion by numbers: protein traffic in prokaryotes.
\newblock {\em Mol Microbiol}, 62:308--319, 2006.

\bibitem{Palmer2012}
T~Palmer and B~C Berks.
\newblock The twin-arginine translocation ({T}at) protein export pathway.
\newblock {\em Nat Rev Microbiol}, 10:483--496, 2012.

\bibitem{Frobel2012}
J~Fr\"{o}bel, P~Rose, and M~M\"{u}ller.
\newblock Twin-arginine-dependent translocation of folded proteins.
\newblock {\em Philos Trans R Soc London, Ser B}, 367:1029--1046, 2012.

\bibitem{DeBuck2008}
E~De Buck, E~Lammertyn, and J~Anne.
\newblock The importance of the twin-arginine translocation pathway for
  bacterial virulence.
\newblock {\em Trends Microbiol}, 16:442--453, 2008.

\bibitem{Bruser2007}
T~Br\"{u}ser.
\newblock The twin-arginine translocation system and its capability for protein
  secretion in biotechnological protein production.
\newblock {\em Appl Microbiol Biotechnol}, 76:35--45, 2007.

\bibitem{Leake2008}
M~C Leake, N~P Greene, R~M Godun, T~Granjon, G~Buchanan, S~Chen, R~M Berry,
  T~Palmer, and B~C Berks.
\newblock Variable stoichiometry of the {T}at{A} component of the twin-arginine
  protein transport system observed by \emph{in vivo} single-molecule imaging.
\newblock {\em Proc Natl Acad Sci USA}, 105:15376--15381, 2008.

\bibitem{Gohlke2005}
U~Gohlke, L~Pullan, C~A McDevitt, I~Porcelli, E~de~Leeuw, T~Palmer, H~R Saibil,
  and B~C Berks.
\newblock The {T}at{A} component of the twin-arginine protein transport system
  forms channel complexes of variable diameter.
\newblock {\em Proc Natl Acad Sci USA}, 102:10482--10486, 2005.

\bibitem{Berks2000}
B~C Berks, F~Sargent, and T~Palmer.
\newblock The {T}at protein export pathway.
\newblock {\em Mol Microbiol}, 35:260--274, 2000.

\bibitem{DeLisa2003}
M~P DeLisa, D~Tullman, and G~Georgiou.
\newblock Folding quality control in the export of proteins by the bacterial
  twin-arginine translocation pathway.
\newblock {\em Proc Natl Acad Sci USA}, 100:6115--6120, 2003.

\bibitem{Matos2008}
C~F R~O Matos, C~Robinson, and A~Di Cola.
\newblock The {T}at system proofreads {F}e{S} protein substrates and directly
  initiates the disposal of rejected molecules.
\newblock {\em EMBO J}, 27:2055--2063, 2008.

\bibitem{Panahandeh2008}
S~Panahandeh, C~Maurer, M~Moser, M~P DeLisa, and M~M\"{u}ller.
\newblock Following the path of a twin-arginine percursor along the {T}at{ABC}
  translocase of \emph{{E}scherichia coli}.
\newblock {\em J Biol Chem}, 283:33267--33275, 2008.

\bibitem{Maurer2009}
C~Maurer, S~Panahandeh, M~Moser, and M~M\"{u}ller.
\newblock Impairment of twin-arginine-dependent export by seemingly small
  alterations of substrate conformation.
\newblock {\em FEBS Lett}, 583:2849--2853, 2009.

\bibitem{Cline2007}
K~Cline and M~McCaffery.
\newblock Evidence for a dynamic and transient pathway through the {TAT}
  protein transport machinery.
\newblock {\em EMBO J}, 26:3039--3049, 2007.

\bibitem{Richter2007}
S~Richter, U~Lindenstrauss, C~Lucke, R~Bayliss, and T~Br\"{u}ser.
\newblock Functional {T}at transport of unstructured, small, hydrophilic
  proteins.
\newblock {\em J Biol Chem}, 282:33257--33264, 2007.

\bibitem{Rocco2012}
M~A Rocco, D~Waraho-Zhmayev, and M~P DeLisa.
\newblock Twin-arginine translocase mutations that suppress folding quality
  control and permit export of misfolded substrate proteins.
\newblock {\em Proc Natl Acad Sci USA}, 109:13392--13397, 2012.

\bibitem{Palmer2005}
T~Palmer, F~Sargent, and B~C Berks.
\newblock Export of complex cofactor-containing proteins by the bacterial {T}at
  pathway.
\newblock {\em Trends Microbiol}, 13:175--180, 2005.

\bibitem{Lindenstrauss2009}
U~Lindenstrau{\ss} and T~Br\"{u}ser.
\newblock Tat transport of linker-containing proteins in \emph{{E}scherichia
  coli}.
\newblock {\em FEMS Microbiol Lett}, 295:135--140, 2009.

\bibitem{Richter2005}
S~Richter and T~Br\"{u}ser.
\newblock Targeting of unfolded {P}ho{A} to the {TAT} translocon of
  \emph{{E}scherichia coli}.
\newblock {\em J Biol Chem}, 280:42723--42730, 2005.

\bibitem{Whitaker2012}
N~Whitaker, U~K Bageshwar, and S~M Musser.
\newblock Kinetics of precursor interactions with the bacterial {T}at
  translocase detected by real-time {FRET}.
\newblock {\em J Biol Chem}, 287:11252--11260, 2012.

\bibitem{Musser2000}
S~M Musser and S~M Theg.
\newblock Characterization of the early steps of {OE}17 precursor transport by
  the thylakoid {$\Delta$}p{H}/{T}at machinery.
\newblock {\em Eur J Biochem}, 267:2588--2598, 2000.

\bibitem{Bageshwar2009}
U~K Bageshwar, N~Whitaker, F-C Liang, and S~M Musser.
\newblock Interconvertibility of lipid- and translocon-bound forms of the
  bacterial {T}at precursor pre-{S}ufl.
\newblock {\em Mol Microbiol}, 74:209--226, 2009.

\bibitem{Lindenstrauss2010}
U~Lindenstrau{\ss}, C~F R~O Matos, W~Graubner, C~Robinson, and T~Br\"{u}ser.
\newblock Malfolded recombinant {T}at substrates are {T}at-independently
  degraded in \emph{{E}scherichia coli}.
\newblock {\em FEBS Lett}, 584:3644--3648, 2010.

\bibitem{Alcock2013}
F~Alcock, M~A~B Baker, N~P Greene, T~Palmer, M~I Wallace, and B~C Berks.
\newblock Live cell imaging shows reversible assembly of the {T}at{A} component
  of the twin-arginine protein transport system.
\newblock {\em Proc Natl Acad Sci USA}, 110:E3650--E3659, 2013.

\bibitem{Tarry2009}
M~J Tarry, E~Sch\"{a}fer, S~Chen, G~Buchanan, N~P Greene, S~M Lea, T~Palmer,
  H~R Saibil, and B~C Berks.
\newblock Structural analysis of substrate binding by the {T}at{BC} component
  of the twin-arginine protein transport system.
\newblock {\em Proc Natl Acad Sci USA}, 106:13284--13289, 2009.

\bibitem{Stanley2000}
N~R Stanley, T~Palmer, and B~C Berks.
\newblock The twin arginine consensus motif of {T}at signal peptides is
  involved in {S}ec-independent protein targeting in \emph{{E}scherichia coli}.
\newblock {\em J Biol Chem}, 275:11591--11596, 2000.

\bibitem{Hinsley2001}
A~P Hinsley, N~R Stanley, T~Palmer, and B~C Berks.
\newblock A naturally occurring bacterial {T}at signal peptide lacking one of
  the `invariant' arginine residues of the consensus targeting motif.
\newblock {\em FEBS Lett}, 497:45--49, 2001.

\bibitem{ma10}
X~Ma and K~Cline.
\newblock Multiple precursor proteins bind individual {T}at receptor complexes
  and are collectively transported.
\newblock {\em EMBO J}, 29:1477--1488, 2010.

\bibitem{Kostecki2010}
J~S Kostecki, H~Li, R~J Turner, and M~P DeLissa.
\newblock Visualizing interactions along the \emph{{E}scherichia coli}
  twin-arginine translocation pathway using protein fragment complementation.
\newblock {\em PLOS One}, 5:e9225, 2010.

\bibitem{Gillespie1977}
D~T Gillespie.
\newblock Exact stochastic simulation of coupled chemical reactions.
\newblock {\em J Phys Chem}, 81:2340--2361, 1977.

\bibitem{Zilman2010}
A~Zilman, S~Di Talia, T~Jovanovic-Talisman, B~T Chait, M~P Rout, and M~O
  Magnasco.
\newblock Enhancement of transport selectivity through nano-channels by
  non-specific competition.
\newblock {\em PLoS Comput Biol}, 6:e1000804, 2010.

\bibitem{Berg1985}
O~G Berg and P~H von Hippel.
\newblock Diffusion-controlled macromolecular interactions.
\newblock {\em Ann Rev Biophys Biophys Chem}, 14:131--160, 1985.

\bibitem{Howan2012}
K~Howan, A~J Smith, L~F Westblade, N~Joly, W~Grange, S~Zorman, S~A Darst, N~J
  Savery, and T~R Strick.
\newblock Initiation of transcription-coupled repair characterized at
  single-molecule resolution.
\newblock {\em Nature}, 490:431--434, 2012.

\bibitem{DabneySmith2006}
C~Dabney-Smith, H~Mori, and K~Cline.
\newblock Oligomers of {T}ha4 organize at the thylakoid {T}at translocase
  during protein transport.
\newblock {\em J Biol Chem}, 281:5476--5483, 2006.

\bibitem{Whitaker2013}
N~Whitaker, U~Bageshwar, and S~M Musser.
\newblock Effect of cargo size and shape on the transport efficiency of the
  bacterial {T}at translocase.
\newblock {\em FEBS Lett}, 587:912--916, 2013.

\bibitem{Muller2005}
M~M\"{u}ller and R~B Kl\"{o}sgen.
\newblock The {T}at pathway in bacteria and chloroplasts.
\newblock {\em Mol Membr Biol}, 22:113--121, 2005.

\bibitem{Alder2003}
N~N Alder and S~M Theg.
\newblock Energetics of protein transport across biological membranes: a study
  of the thylakoid ${\Delta}$p{H}-dependent/cp{T}at pathway.
\newblock {\em Cell}, 112:231--242, 2003.

\bibitem{Asai1999}
T~Asai, Y~Shinoda, T~Nohara, T~Yoshihisa, and T~Endo.
\newblock Sec-dependent pathway and ${\Delta}$p{H}-dependent pathway do not
  share a common translocation pore in thylakoidal protein transport.
\newblock {\em J Biol Chem}, 274:20075--20078, 1999.

\bibitem{Antal2013}
T~K Antal, I~B Kovalenko, A~B Rubin, and E~Tyystj\"{a}rvi.
\newblock Photosynthesis-related quantities for education and modeling.
\newblock {\em Photosynth Res}, 117:1--30, 2013.

\bibitem{Jack2004}
R~L Jack, G~Buchanan, A~Dubini, K~Hatzixanthis, T~Palmer, and F~Sargent.
\newblock Coordinating assembly and export of complex bacterial proteins.
\newblock {\em EMBO J}, 23:3962--3972, 2004.

\bibitem{Robinson2011}
C~Robinson, C~F R~O Matos, D~Beck, C~Ren, J~Lawrence, N~Vasisht, and S~Mendel.
\newblock Transport and proofreading of proteins by the twin-arginine
  translocation ({T}at) system in bacteria.
\newblock {\em Biochim Biophys Acta}, 1808:876--884, 2011.

\end{thebibliography}
\end{document}


\title{Supplemental: Tat translocation without specific quality control}
\author{Chitra R Nayak, Aidan I Brown, and Andrew D Rutenberg }



\begin{figure}[h]  
 \begin{center}
\includegraphics[width=15cm]{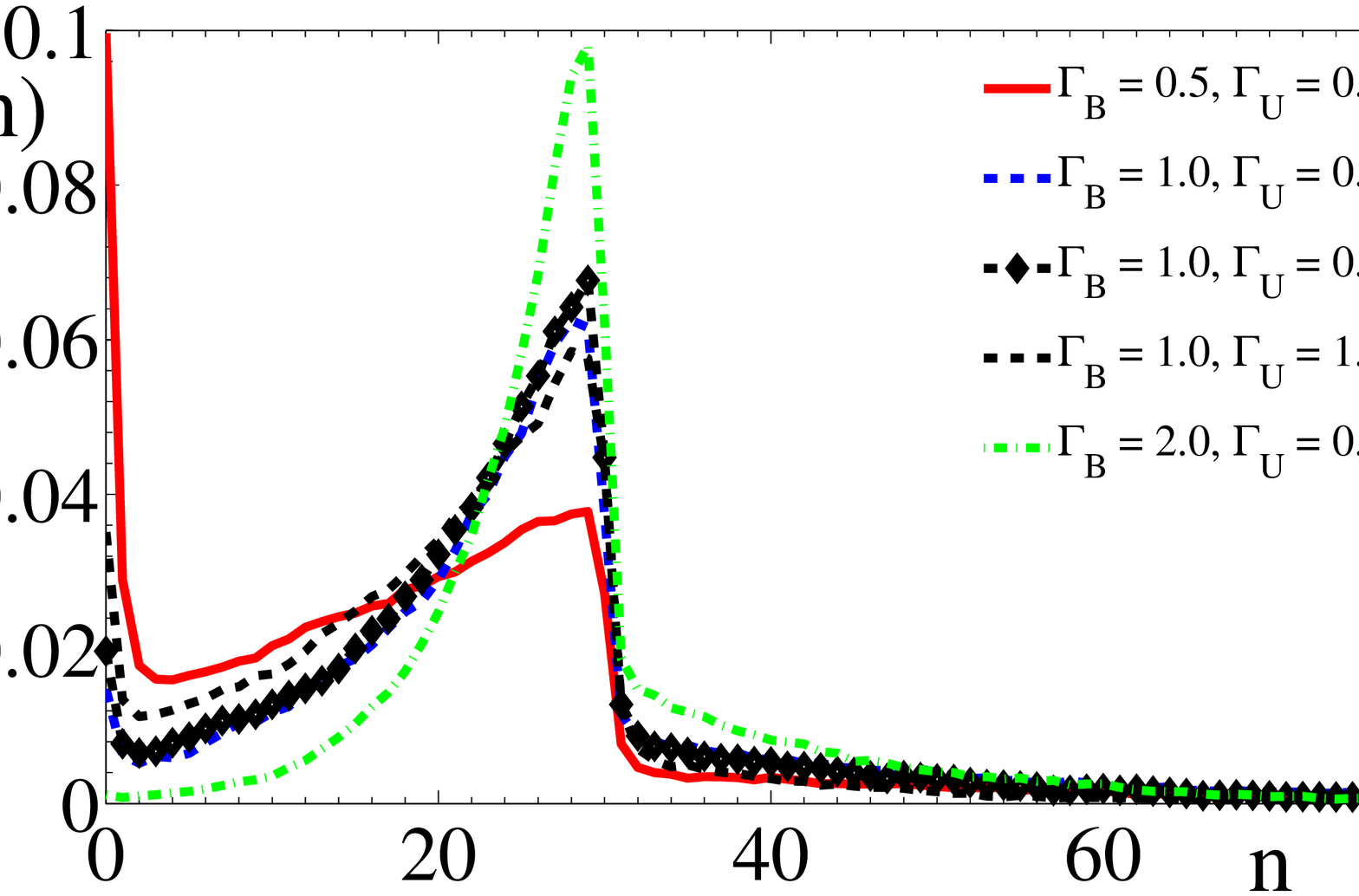}
  \end{center}
\caption{{\bf Effect of $\Gamma_B$ and  $\Gamma_U.$} The cluster size distribution $P(n)$ vs.\ the cluster size $n$ as the binding rate $\Gamma_B$ and the unbinding rate $\Gamma_U$ are varied. All of the other parameters are at their default values. The unbinding rate does not make dramatic changes to the cluster size distribution. In contrast, the binding rate significantly changes the distribution. A smaller binding rate gives more time for clusters without substrates to shrink, diminishing the tail at $n>n_c$ and increasing the weight at small clusters $n<n_c$. All rates other than those in the legend are the same as in Fig.~2, with $f=0.1$. }
\label{Fig:gamBgamU}
\end{figure}

\begin{figure}[h] 
 \begin{center}
\includegraphics[width=15cm]{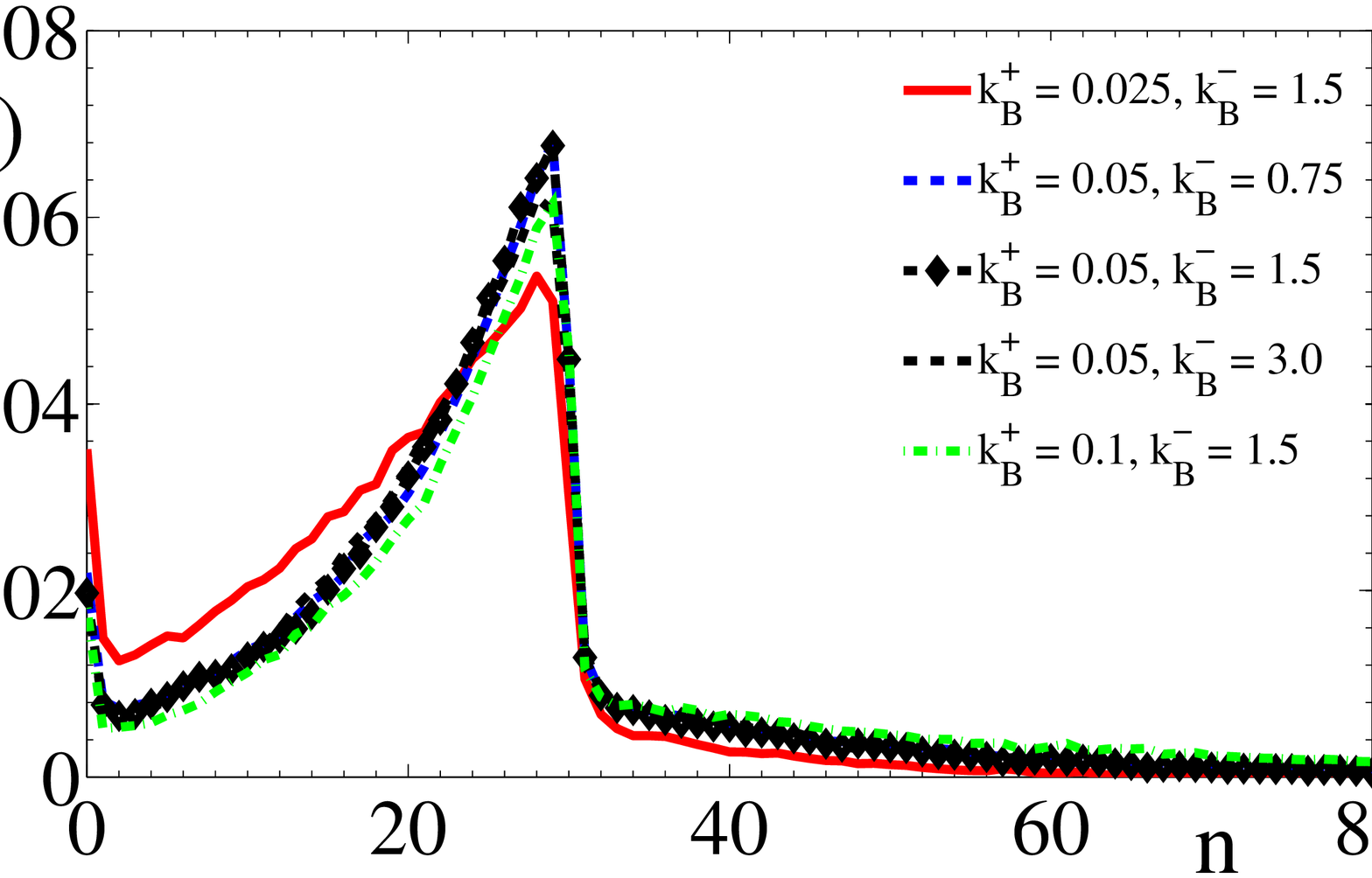}
  \end{center}
\caption{{\bf Effect of $k_B^+$ and  $k_B^-.$} The cluster size distribution $P(n)$ vs.\ the cluster size $n$ as the bound cluster growth and decay rates are varied. The changes are modest, but the $n>n_c$ tail is increased with increasing $k_B^+$ and with decreasing $k_B^-$. All rates other than those in the legend are the same as in Fig.~2, with $f=0.1$.}
\label{Fig:kbgkbd}
\end{figure}

\begin{figure}[h] 
 \begin{center}
\includegraphics[width=15cm]{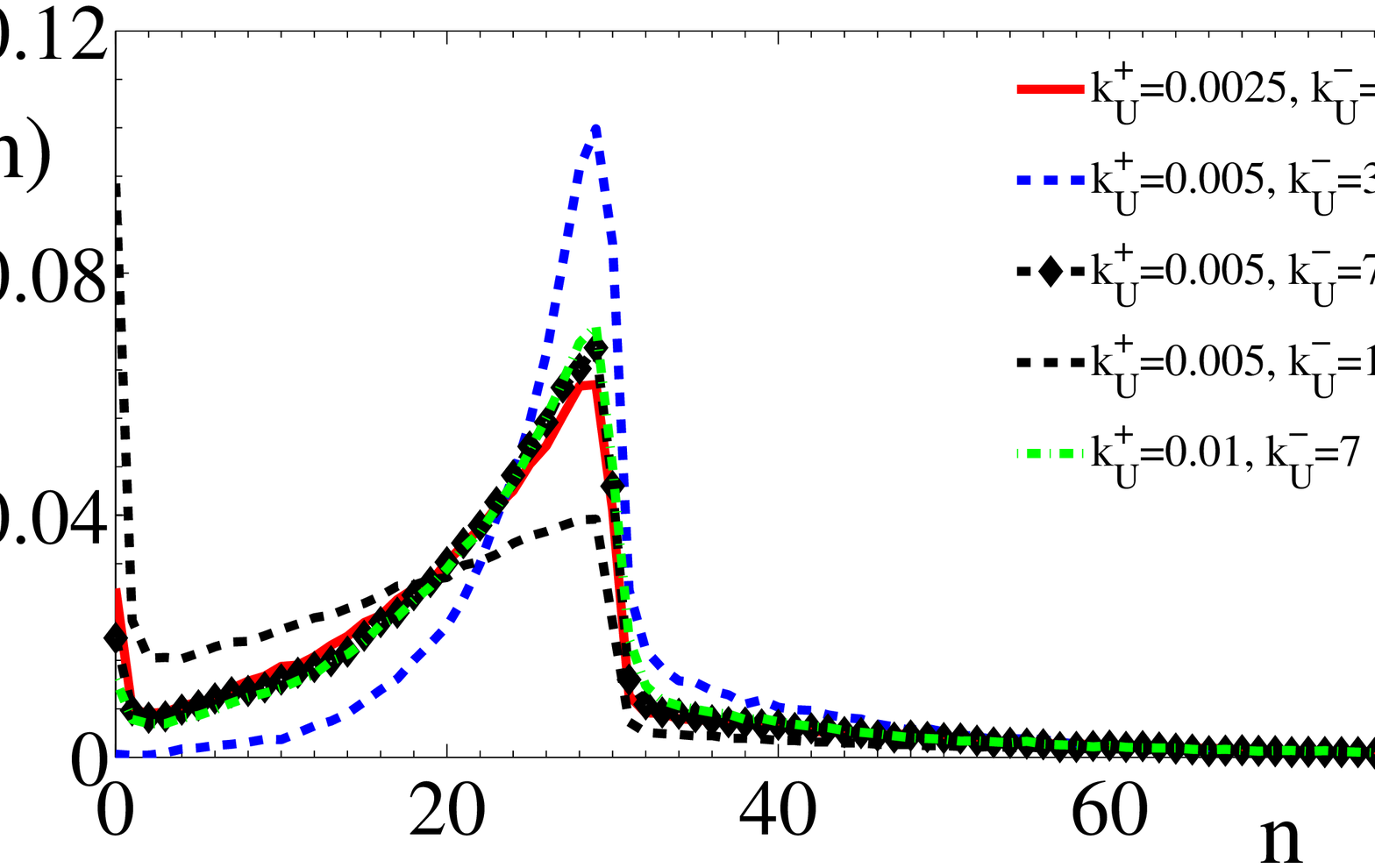}
  \end{center}
\caption{{\bf Effect of $k_U^+$ and  $k_U^-.$} The cluster size distribution $P(n)$ vs.\ the cluster size $n$ as the unbound cluster growth and decay rates are varied. Unbound decay dominates the kinetics, and changes to $k_U^-$ strongly affect the distribution. Increasing $k_U^-$ decreases the tail for $n>n_c$ and significantly increases the weight below $n_c$ as well.  Changes due to variations of $k_U^+$ are relatively small. All rates other than those in the legend are the same as in Fig.~2, with $f=0.1$.}
\label{Fig:kugkud}
\end{figure}